\documentclass[12pt]{article}

\usepackage{graphicx}

\usepackage{latexsym} 
\usepackage{amssymb}  

\newcommand{\al}{\alpha}
\newcommand{\be}{\beta}
\newcommand{\ga}{\gamma}

\newcommand{\de}{\delta}
\newcommand{\De}{\Delta}

\newcommand{\ka}{\kappa}

\newcommand{\La}{\Lambda}

\newcommand{\<}{\langle} 
\renewcommand{\>}{\rangle} 

\newcommand{\dsp}{\displaystyle}
\newcommand{\ad}{\dagger}
\newcommand{\beq}{\begin{equation}}
\newcommand{\eeq}{\end{equation}}
\newcommand{\ba}{\begin{array}}
\newcommand{\bea}{\begin{eqnarray}}
\newcommand{\ea}{\end{array}}
\newcommand{\eea}{\end{eqnarray}}
\newcommand{\ben}{\begin{enumerate}} 
\newcommand{\een}{\end{enumerate}}
\newcommand{\bc}{\begin{center}}
\newcommand{\ec}{\end{center}}
\newcommand\hide[1]{}

\renewcommand{\O}{{\cal O}}

\newcommand{\Dash}{\boldmath $-$}  
\newcommand{\skipover}[1]{}

\newcommand{\km}{{\rm km}} 
\newcommand{\fm}{{\rm fm}} 
 
\newcommand{\keV}{{\rm keV}} 
\newcommand{\MeV}{{\rm MeV}} 
 
\newcommand{\Qt}{{\tilde Q} }

\newcommand{\Msolar}{M_\odot}

\newcommand{\nsat}{n_{\rm sat}} 

\begin{document}

\title{Dense quark matter in compact stars}

\author{M. Alford \\[1ex]
Physics Department \\ Washington University CB 1105 \\
Saint Louis, MO 63130 \\ USA
}

\date{29 May 2003}

\begin{titlepage}
\maketitle

\begin{abstract}
The densest predicted state of matter is
colour-superconducting quark matter, in which quarks
near the Fermi surface form a condensate of Cooper pairs.
This form of matter may well exist in the core of compact
stars, and the search for signatures of its presence is an
ongoing enterprise.
Using a bag model of quark matter,
I discuss the effects of colour superconductivity on the
mass-radius relationship of compact stars, showing that
colour superconducting quark matter can occur in
compact stars at values of the bag constant where ordinary quark
matter would not be allowed.  The resultant ``hybrid''
stars with colour superconducting quark matter interior and nuclear
matter surface have masses in the range $1.3$-$1.6~\Msolar$ and radii
$8$-$11~\km$. Once perturbative corrections are included,
quark matter can show a mass-radius relationship very similar to that
of nuclear matter, and the mass of a hybrid star can reach $1.8~\Msolar$.
\end{abstract}

\end{titlepage}

\section{Introduction}
\label{sec:intro}

It has become clear over the last few years that the phase diagram
of QCD is much richer than originally
believed. In addition to the hadronic phase with which we are
familiar and the quark gluon plasma (QGP) that is predicted
to lie at temperatures above $170~\MeV$, there is a
whole family of ``colour superconducting'' phases, that are expected
to occur at high density and low temperature \cite{Reviews}.
The essence of colour superconductivity is quark pairing, driven by
the BCS mechanism, which operates when there exists
an attractive interaction between fermions at a Fermi surface.
The QCD quark-quark interaction is strong, and is attractive
in many channels, so we expect cold dense quark matter to {\em generically}
exhibit colour superconductivity.
Moreover, quarks, unlike electrons, have colour and flavour as well as spin
degrees of freedom, so many different patterns of pairing are possible.
This leads us to expect a rich phase structure
in matter beyond nuclear density.

Colour superconducting quark matter may occur naturally in the
universe, in the cold dense cores of neutron stars.
(It may also be possible to create it
in low-energy heavy ion colliders, such as
the proposed Compressed Baryonic Matter facility at GSI Darmstadt.)
There densities are above nuclear
density, and temperatures are of the order of tens of \keV. Thus most
work on signatures has focussed on properties of colour
superconducting quark matter that would affect observable features of
compact stars, and I will discuss some of these below.

\section{Phase diagram of QCD}

\begin{figure}[t]
\parbox{0.48\hsize}{
\bc \underline{Heavy strange quark} \ec
 \includegraphics[width=\hsize]{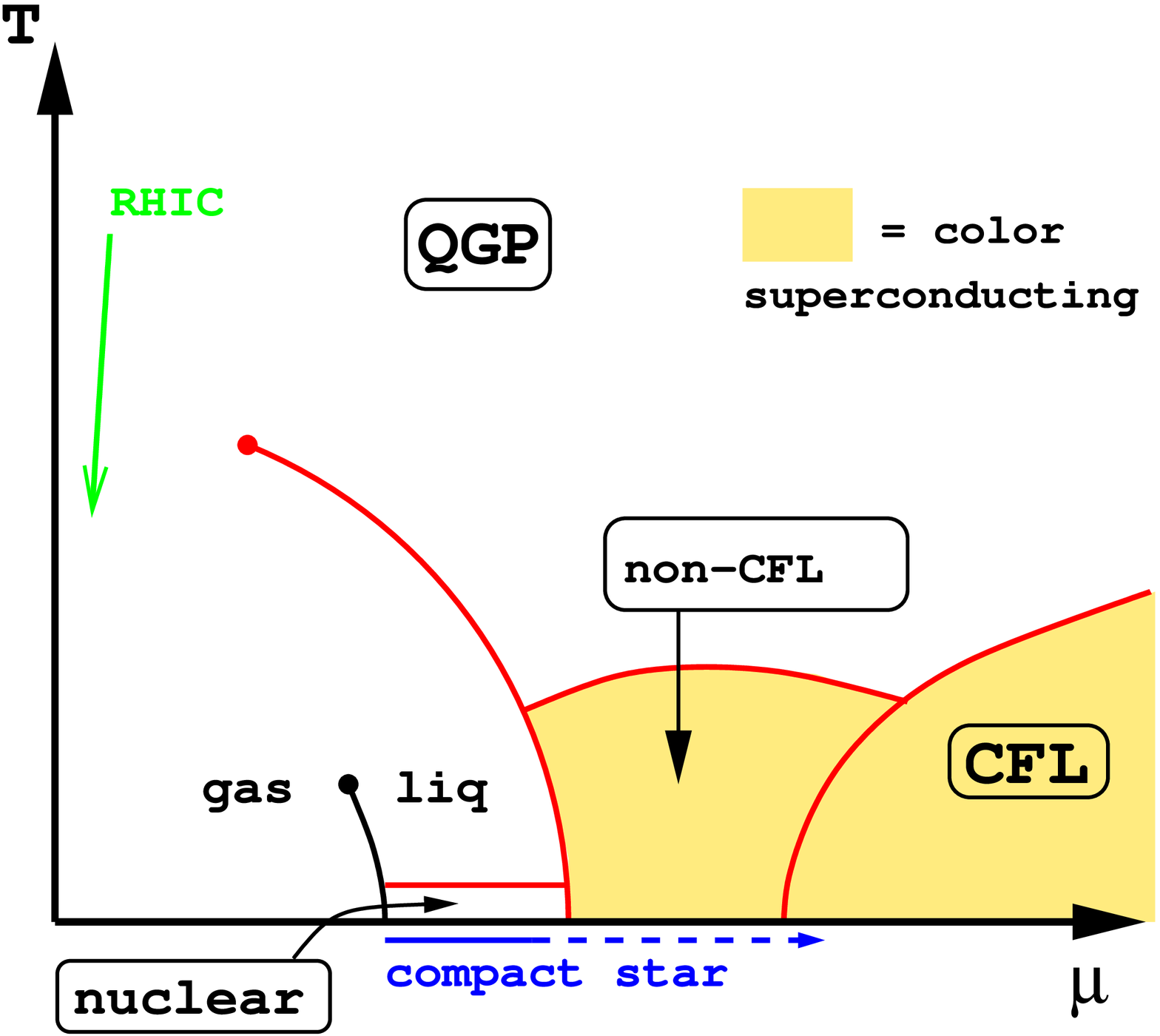}
}
\parbox{0.48\hsize}{
\bc \underline{Light strange quark} \ec
 \includegraphics[width=\hsize]{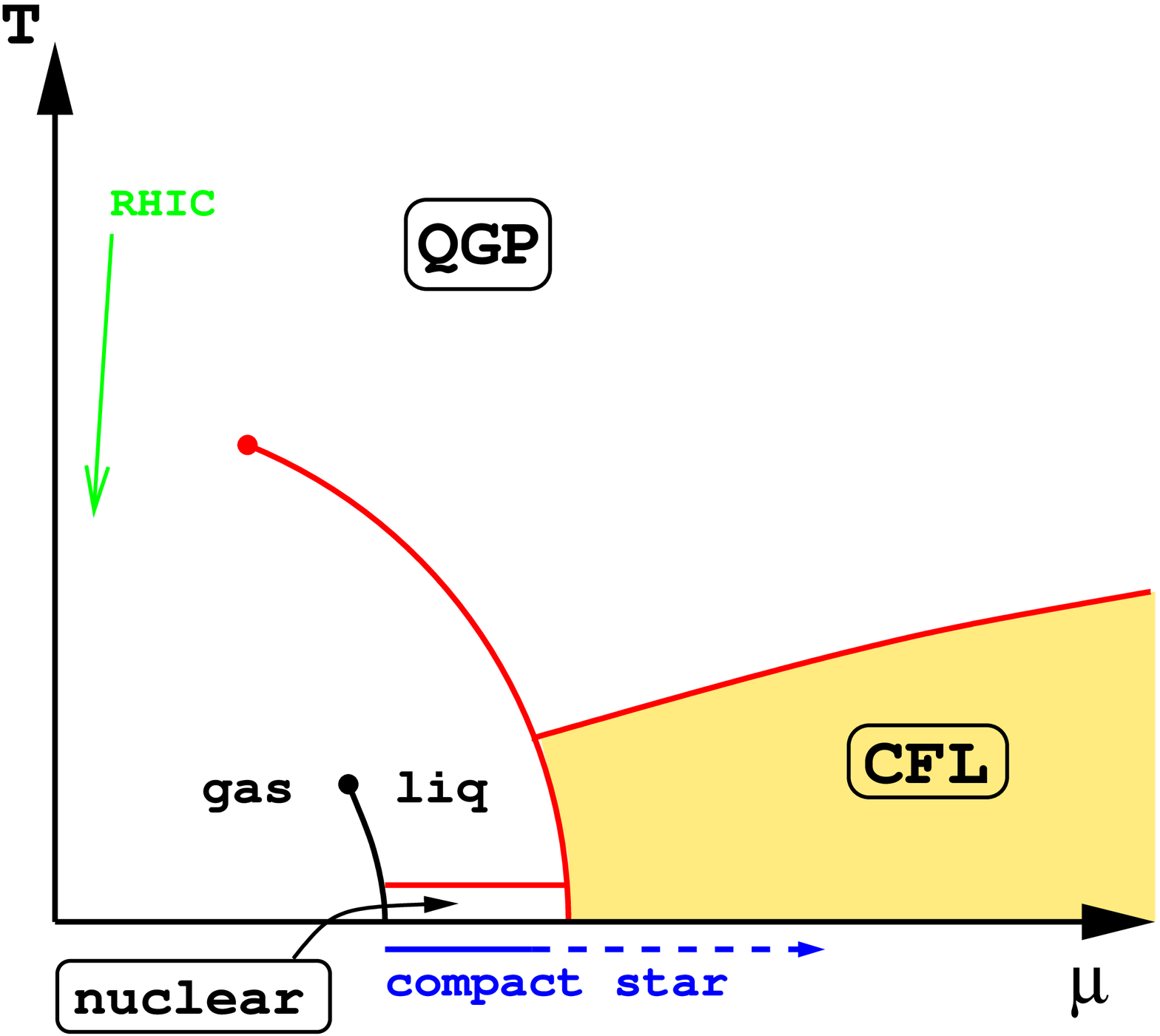}
}
\caption{Conjectured phase diagrams for QCD in the real world.
For a light strange quark, there is a direct transition from
nuclar matter to colour-flavour locked colour
superconducting quark matter. For a heavy strange quark,
there is an intermediate phase where the strange quark
pairs in some other way. Depending on the strength of instanton 
interactions, the CFL phase may include $K^0$ condensation.}
\label{fig:phase}
\end{figure}

The strange quark plays a crucial role in the phases of QCD.
In Fig.~\ref{fig:phase} I show two conjectured phase diagrams for QCD,
one for a light strange quark and one for a heavy strange quark.
In both cases, along the horizontal axis the temperature is zero, and
the density rises from the onset of nuclear matter through the
transition to
quark matter. Compact stars are in this region of the phase diagram,
although it is not known whether their cores are dense enough
to reach the quark matter phase.
Along the vertical axis the temperature rises, taking
us through the crossover from a hadronic gas to the quark gluon plasma.
This is the regime explored by high-energy heavy-ion colliders
(see other contributions to these proceedings).

If the effective strange quark mass is low enough in quark matter
at a few times nuclear density, then there is a direct transition
from nuclear matter to colour-flavour-locked (CFL) quark matter \cite{ARW3}. 
In the CFL phase the strange
quark participates symmetrically with the up and down quarks
in Cooper pairing---this is described
in more detail below. If the strange quark is too heavy to pair
symmetrically with the light quarks at these densities, 
then there will be an interval of some other phase.
This may be crystalline colour superconductivity \cite{OurLOFF}
or some form of single-flavour pairing \cite{oneflav}.

\section{Colour superconductivity}

To understand the essence of colour superconductivity,
we can borrow some intuition from
condensed matter physics
\cite{oldcolorSC,newcolorSC}.
QCD is asymptotically free---the interaction becomes weaker as the
momentum grows---so at sufficiently high density and low temperature,
there is a  Fermi surface of almost free quarks. 
The interactions between
quarks near the Fermi surface are certainly attractive in some channels
(quarks bind together to form baryons)
and it was shown by Bardeen, Cooper, and
Schrieffer (BCS) \cite{BCS} that if there is {\em any} channel in which the
interaction is attractive, then there is a state
of lower free energy than a simple Fermi surface. That state arises
from  a complicated coherent 
superposition of pairs of particles (and holes)---``Cooper pairs''.

We can understand the BCS mechanism in an intuitive way as follows.
The Helmholtz free energy is $F= E-\mu N$, where $E$ is
the total energy of the system, $\mu$ is the chemical potential, and
$N$ is the number of fermions. The Fermi surface is defined by a
Fermi energy $E_F=\mu$, at which the free energy is minimized, so
adding or subtracting a single particle costs zero free energy. 
Now switch on a weak attractive interaction.
It costs no free energy to
add a pair of particles (or holes), and the attractive
interaction between them then lowers the free energy of the system.
Many such pairs will therefore
be created in the modes near the Fermi surface, and these pairs,
being bosonic, will form a condensate. The ground state will be a
superposition of states with all numbers of pairs, breaking the
fermion number symmetry. 

Since pairs of quarks cannot be colour singlets,
the resulting condensate will break the local colour symmetry
$SU(3)_{\rm colour}$.  We call this ``colour superconductivity''.
Note that the quark pairs play the same role here as the Higgs particle
does in the standard model: the colour-superconducting phase
can be thought of as the Higgs phase of QCD.
For more detailed discussions of calculations of the favored
pairing patterns, see the reviews \cite{Reviews}.

\subsection{Three flavours: Colour-flavour locking (CFL)}
\label{sec:CFL}

The favored pairing pattern at high densities, where
the strange quark Fermi momentum is close to the up and down
quark Fermi momenta, is ``colour-flavour locking'' (CFL). 
In QCD with three flavours of massless quarks 
the Cooper pairs {\em cannot}
be flavour singlets, and both colour and flavour symmetries are
necessarily broken \cite{ARW3}.
Both NJL \cite{ARW3,SW-cont,HsuCFL} and gluon-mediated interaction calculations
\cite{3flavpert} agree that the attractive channel 
exhibits a pattern
called colour-flavour locking (CFL),
\beq
\ba{c}
\< q^\al_i q^\be_j \>^{\phantom\ad}_{1PI}
\propto C \ga_5 \Bigl(
 (\ka+1)\de^\al_i\de^\be_j + (\ka-1) \de^\al_j\de^\be_i \Bigr)
  \\[2ex]
 {[SU(3)_{\rm colour}]}
 \times \underbrace{SU(3)_L \times SU(3)_R}_{\dsp\supset [U(1)_Q]}
 \times U(1)_B 
 \to \underbrace{SU(3)_{C+L+R}}_{\dsp\supset [U(1)_{\Qt}]} 
  \times \mathbb{Z}_2
\ea
\eeq
Colour indices $\al,\be$ and flavour indices $i,j$ run from 1 to 3,
Dirac indices are suppressed,
and $C$ is the Dirac charge-conjugation matrix.
The term multiplied by $\ka$ corresponds to pairing in the
$({\bf 6}_S,{\bf 6}_S)$, which
although not energetically favored
breaks no additional symmetries and so
$\ka$ is in general small but not zero 
\cite{ARW3,PisarskiCFL,3flavpert}.
The Kronecker deltas connect
colour indices with flavour indices, so that the condensate is not
invariant under colour rotations, nor under flavour rotations,
but only under simultaneous, equal and opposite, colour and flavour
rotations. Since colour is only a vector symmetry, this
condensate is only invariant under vector flavour+colour rotations, and
breaks chiral symmetry. The effect of including a strange quark mass
is to distort the CFL pairing \cite{ABR2+1,SW-cont}, and a sufficiently
large strange quark mass will unpair the strange quark completely
\cite{ABR2+1,SW-cont,AR-02}, 
yielding a two-flavour colour superconductor (2SC) \cite{newcolorSC},
and at lower
values it may induce a flavour rotation of the condensate
known as ``kaon condensation'' \cite{BedaqueSchaefer}.
The features of the CFL pattern of condensation are
\begin{itemize}
\setlength{\itemsep}{-0.7\parsep}
\item[\Dash] The colour gauge group is completely broken. All eight gluons
become massive. This ensures that there are no infrared divergences
associated with gluon propagators.
\item[\Dash]
All the quark modes are gapped. The nine quasiquarks 
(three colours times three flavours) fall into an ${\bf 8} \oplus {\bf 1}$
of the unbroken global $SU(3)$, so there are two
gap parameters. The singlet has a larger gap than the octet.
\item[\Dash] 
A rotated electromagnetism (``$\Qt$'')
survives unbroken. It is a combination
of the original photon and one of the gluons.
\item[\Dash] Two global symmetries are broken,
the chiral symmetry and baryon number, so there are two 
gauge-invariant order parameters
that distinguish the CFL phase from the QGP,
and corresponding Goldstone bosons which are long-wavelength
disturbances of the order parameter. 
When the light quark mass is non-zero it explicitly breaks
the chiral symmetry and gives a mass
to the chiral Goldstone octet, but the CFL phase is still
a superfluid, distinguished by its baryon number breaking.
\item[\Dash]
The symmetries of the
3-flavour CFL phase are the same as those one might expect for 3-flavour
hypernuclear matter \cite{SW-cont}, so it is possible that there is
no phase transition between them.
\end{itemize}

In a real neutron star 
we must require electromagnetic and colour neutrality
\cite{BaymIida,AR-02}
(ignoring charge-separated phases).
It turns out that this penalizes the 2SC phase relative to the
CFL phase \cite{AR-02}.
The reason is that the CFL phase has already paid most of
the cost of neutrality, since it brings the $u$, $d$, and $s$
Fermi surfaces close together
\cite{AR-02,CFLneutral,Steiner:2002gx}, and the 2SC phase
only pairs  4 of the 9 quark colours and flavours, so it
has much less pair binding energy than the CFL phase.

The arguments made in Ref.~\cite{AR-02} are model-independent, based
on simplified assumptions about the dependence of the constituent
strange quark mass $M_s$ on $\mu$ and expanding the free energy in
powers of $M_s/\mu$.  The NJL calculation of
Ref.~\cite{Steiner:2002gx} handles $M_s\sim\mu$ and includes the
coupling between the chiral condensate and quark condensate gap
equations.  The net result is that once neutrality is imposed, there
is no (or very little) density range in which 2SC is the phase with
the lowest free energy.

\section{Mass-radius relationship for compact stars}

The high density and relatively low temperature required to produce
colour superconducting quark matter may be attained
in compact stars,
also known as ``neutron stars'', since it is often
assumed that they are made primarily of neutrons.
Typical compact stars have masses
close to $1.4 \Msolar$, and are believed to have radii of order 10 km.
(For reviews see Ref.~\cite{nstar,GBOOK}).

Colour superconductivity affects the equation of state
at order $(\De/\mu)^2$. It also gives mass to
excitations around the ground state: it opens
up a gap at the quark Fermi surface, and makes the gluons
massive. One would therefore expect it to have a profound effect
on transport properties, such as mean free paths,
conductivities and viscosities.

Although the effects of colour superconductivity on the quark matter
equation of state are subdominant, they may have a large effect
on the mass-radius relationship. The reason for this is that
the pressure of quark matter relative to the hadronic vacuum
contains a constant (the ``bag constant'' $B$) that represents
the cost of dismantling the chirally broken and confining
hadronic vacuum,
\beq
p \sim \mu^4 + \De^2\mu^2 - B~.
\label{roughEoS}
\eeq
If the bag constant is large enough so that nuclear matter is favored
(or almost favored) over quark matter at $\mu\sim 320~\MeV$, then the
bag constant and $\mu^4$ terms almost cancel, and the superconducting
gap $\De$ may have a large effect on the equation of state and hence
on the mass-radius relationship of a compact star \cite{Lugones:2002ak}.

\begin{figure}[htb]
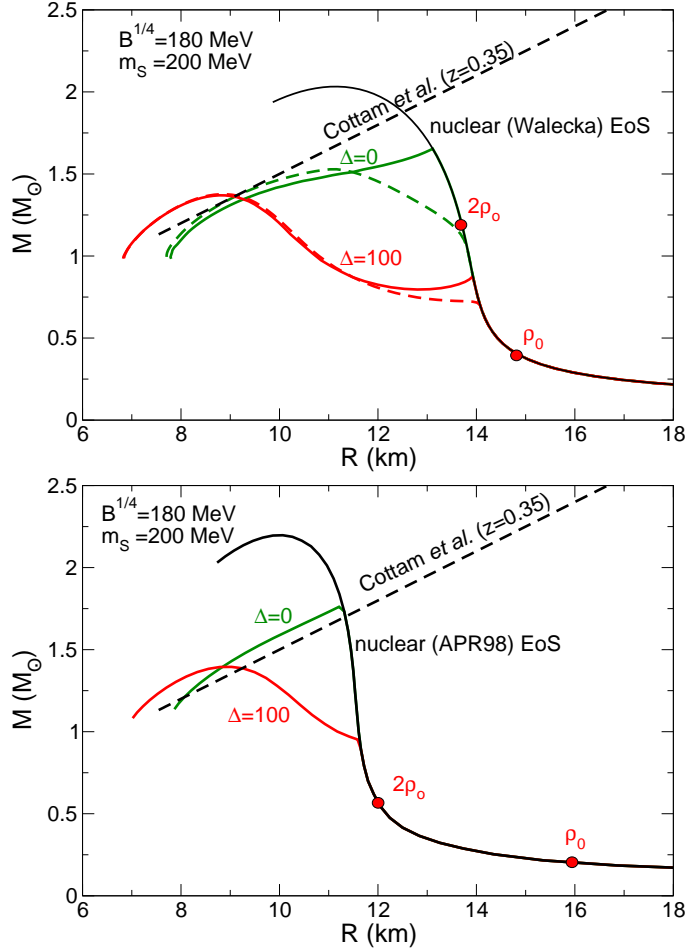

\begin{center}
\includegraphics[width=0.6\textwidth]{figs/mass-radius-wal.eps}
\includegraphics[width=0.6\textwidth]{figs/mass-radius-apr.eps}
\end{center}
\caption{Mass-radius relationships at fixed  bag constant
$B^{1/4}=180~\MeV$ and $m_s=200$ MeV, for unpaired ($\De=0$) 
and colour-superconducting ($\De=100~\MeV$) quark matter. 
The mixed phase (dashed) and the sharp
interface curves are shown. The line labeled ``Cottam {\it et al.}''
indicates the constraint obtained by recent measurements of the
redshift on three spectral lines from EXO0748-676 \cite{Cottam}. 
The dots labeled
$\rho_0$ and $2\rho_0$ on the nuclear matter mass-radius curve
indicate that the central density at these locations correspond to
nuclear and twice nuclear saturation density respectively. The 
top panel uses the Walecka equation of state for nuclear matter,
and the lower panel uses APR98 (in which case
 we only consider the sharp-interface scenario).}
\label{fig:massradius}
\end{figure}

In Ref.~\cite{AlfordReddy} Sanjay Reddy and I explored the effect of quark
pairing on the $M$-$R$ relationship at values of the bag constant that
are consistent with nuclear phenomenology.  Fig.~\ref{fig:massradius}
shows the mass-radius curve for a plausible model of dense matter: the
Walecka nuclear equation of state, 
and quark matter with physically reasonable values of
the bag constant $B^{1/4}=180$ MeV 
($B=137~\MeV\!/\fm^3$) and
strange quark mass $m_s=200$ MeV \cite{GBOOK}. 
Curves for unpaired ($\De=0$) and colour-superconducting ($\De=100~\MeV$)
quark matter are shown. At these  values
the stars are typically ``hybrid'',
containing both quark matter and nuclear matter.  The solid lines in
Fig.~\ref{fig:massradius} correspond to stars that either have no QM
at all, or a sharp transition between NM and QM: the core is made of
quark matter, which is the favored phase at high pressure, and at some
radius there is a transition to nuclear matter, which is favored at
low pressure.  The transition pressure is sensitive to $\De$, for
reasons discussed earlier. The dashed lines are for stars that contain
a mixed NM-QM phase.  In all cases we see that light, large stars
consist entirely of nuclear matter. When the star becomes heavy
enough, the central pressure rises to a level where QM, either in
a mixed phase or in its pure form, occurs in the core. As can be seen
from the figure the transition density is very sensitive to $\De$.

\clearpage

In Ref.~\cite{AlfordReddy} we kept the bag constant fixed,
assuming that it could be fixed by other observations, and 
we treated the quark matter as free quarks with a pairing
energy. It is interesting to see what happens when we relax 
these assumptions.
To see how closely quark matter can mimic nuclear matter, 
we tune the bag constant to keep constant the density of
nuclear matter at its transition to quark matter.
To allow for effects of quark interactions beyond Cooper pairing,
we follow the parameterization of Fraga et.~al.~\cite{Fraga},
who find that the
$\O(\alpha_s^2)$ pressure for three flavours over a finite interval in
$\mu$ can be mimicked by a bag model inspired form given by
\begin{equation}
\ba{rcl}
P_{\alpha_s^2}(\mu_{\rm low} \le \mu \le \mu_{\rm high}) &=& 
\dsp \frac{3}{4\pi^2}~a_{\rm eff}~\mu^4 - B_{\rm eff}\, ,\\[2ex]
a_{\rm eff} &\equiv & 1-c\ .
\ea
\label{peff}
\end{equation}
Choosing the renormalization scale $\Lambda=2\mu$ they find that
$a_{\rm eff}=0.628$ ($c=0.372$) and $B_{\rm eff}^{1/4}=223 $ MeV in the range
$\mu=425$ to $650$ MeV and for $\Lambda=3\mu$ they find that
$a_{\rm eff}=0.626$ ($c=0.374$) and $B_{\rm eff}^{1/4}=157 $ MeV in the range
$\mu=300$ to $470$ MeV. 

\begin{figure}[htb]
\begin{center}
\includegraphics[width=0.7\textwidth,angle=-90]{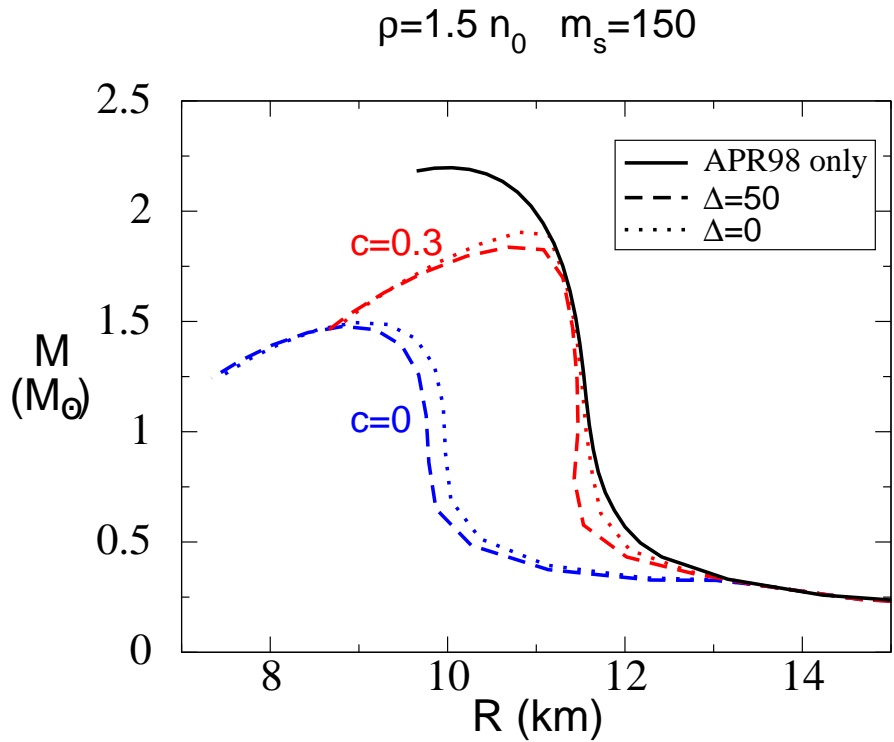}
\end{center}
\caption{$M(R)$ relationship for APR98 nuclear
matter with various quark matter equations of state.
The strange quark is light, and the bag constant is tuned so that
the nuclear matter to quark matter transition occurs at 1.5 times
nuclear saturation density. Dotted lines are unpaired quark matter,
dashed lines are CFL with gap of $50~\MeV$. 
Note how the curve for CFL quark matter with perturbative correction 
but no colour superconductivity ($c=0.3$, $\De=0$; red dotted)
closely follows the pure nuclear curve up to $M\approx 1.8 \Msolar$.
}
\label{fig:rho1.5}
\end{figure}

First, we explore the effect of a colour superconducting gap $\De$
and perturbative correction $c$ on the mass-radius relationship.
We fix the bag constant by requiring that that the 
nuclear to quark matter phase
transition occur at nuclear matter baryon density 
$\rho=1.5 \nsat$. The resultant $M(R)$ curves
are shown in Fig.~\ref{fig:rho1.5}.

\ben
\item The stars resulting from
quark matter equations of state without perturbative correction
($c=0$,blue lines) are smaller and lighter.
This is because they are inherently highly
favored, and so requiring the phase transition to occur at $\rho=1.5 \nsat$
leads to a large bag constant. By contrast, the equations of state with 
perturbative correction ($c=0.4$, red lines) are inherently less favored, 
and require a
low bag constant to bring the phase transition down to $\rho=1.5 \nsat$.
\item
The stars with $c=0.3$ have mass-radius relationships that are very similar to
the pure nuclear APR98 matter.  In fact, for the case where there are
perturbative corrections but no colour superconductivity
the equations of state ($p(\mu)$) are so similar that our program found
a series of phase transitions back and forth between CFL and APR98
up to $\mu=546~\MeV$ (baryon density $\rho=5.4 \nsat$). 
This is why the $c=0.3$ red dotted curve lies almost
exactly on top of the solid black (APR98) curve, even though there
was a phase transition from APR98 to CFL at $\rho=1.5 \nsat$
(which is first attained when the APR98 star reaches a mass of
$0.315 \Msolar$, $R=13.3~\km$). 
\item
In Ref.~\cite{AlfordReddy} we showed that at fixed bag constant,
colour superconductivity has a strong effect on the mass-radius
relationship of compact stars. Here, by comparing the
dashed lines with the dotted lines in Fig.~\ref{fig:rho1.5},
we see that it is difficult to distinguish the effect of
colour superconductivity from a change in the bag constant.
In Fig.~\ref{fig:rho1.5}, as we vary parameters $c$ and $\De$
of the quark matter equation of state, the bag constant 
is tuned to maintain a constant value of the nuclear density
at the transition to quark matter, and in this situation
colour superconductivity only makes a small difference to the
mass-radius relationship.
\een

\section{Other phenomenology}

Mass and size are the most obvious observable properties of
a compact star, but various groups are working on other
possible signatures of colour-superconducting quark matter.

\noindent {\em Interfaces and mixed phases.}
These were studied in
Ref.~\cite{ARRW}, and it was found that a mixed
phase only occurs if the surface tension of the interface
is less than about $40~\MeV/\fm^2 = 0.2\times (200~\MeV)^3$, 
a fairly small value
compared to the relevant scales $\La_{\rm QCD}\approx 200~\MeV$,
$\mu\sim 400~\MeV$. 
A sharp nuclear-quark interface will have an energy-density discontinuity
across it, which will affect gravitational waves emitted in
mergers, and also the $r$-mode spectrum
and the damping forces to which $r$-modes are subject.

\noindent {\em Crystalline pairing (the ``LOFF'' phase)}.
This is expected to occur when two different types of quark have sufficiently
different Fermi momenta
that BCS pairing cannot occur \cite{OurLOFF}.
This is a candidate for the intermediate phase of Fig.~\ref{fig:phase},
where the strange quark mass, combined with requirements of weak equilibrium
and charge neutrality, gives each quark flavour a different Fermi momentum.
The phenomenology of the crystalline phase has not yet been worked out,
but recent calculations using Landau-Ginzburg effective theory 
indicate that the favored phase may be a face-centered cubic crystal
 \cite{Bowers:2002xr}, with a reasonably large binding energy.
This raises the interesting possibility of glitches in quark matter stars.

\noindent {\em Cooling by neutrino emission}.
The cooling rate is
determined by the heat capacity and emissivity, both
of which are sensitive to the spectrum of low-energy excitations,
and hence to colour superconductivity .
CFL quark matter, where all modes are gapped, has a much
smaller neutrino emissivity and heat capacity than nuclear matter, and
hence the cooling of a compact star is likely to be dominated by the
nuclear mantle rather than the CFL core 
\cite{Page,Shovkovy-02,Jaikumar:2002vg}.  
Other phases such as 2SC or LOFF give large gaps to only
some of the quarks. Their cooling would proceed quickly, then
slow down suddenly when the temperature fell below
the smallest of the small weak-channel gaps. This behavior should be
observable.

\noindent {\em $r$-mode spindown}.
The $r$-mode is a
bulk flow in a rotating star that, if the
viscosity is low enough, radiates away energy and angular momentum
in the form of gravitational waves.
One can rule out certain models for compact stars
on the grounds that they have such low damping that 
they could not support the high rotation rates observed in pulsars:
$r$-mode spindown would have slowed them down.
Madsen~\cite{MadsenRmode} has shown that
for a compact star made {\em entirely} of quark matter
in the CFL phase, even a gap as small as $\Delta=1$~MeV
is ruled out by observations of millisecond pulsars.
It remains to extend this calculation to the more generic picture of
a quark matter core surrounded by a nuclear mantle.

\bc{\bf Acknowledgments}\ec
I thank the organizers of SQM 2003. The work reported in Section 4 was
performed in collaboration with Sanjay Reddy, and was supported
by the UK PPARC and by the U.S. Department of Energy under grant number
DE-FG02-91ER40628.


\end{document}